\documentstyle[12pt,moriond,epsf]{article}
\begin{document}
\raisebox{0cm}[0cm][0cm]{\makebox[0cm][l]
{\hspace{-1.5cm}\parbox{14cm}{\em Paper presented at the 
XVIIIth Moriond Astrophysics Meeting: Dwarf Galaxies and Cosmology,
eds. T.X. Thuan, C. Balkowski, V. Cayatte and J. Tran Thanh Van,
Editions Frontieres, Gif-sur-Yvette, 1998 March 14-21
}}}
\heading{THE DISCOVERY OF THE SMALLEST BLUE COMPACT DWARF GALAXIES}

\author{M.J. Drinkwater $^{1}$, M.D. Gregg $^{2,3}$, R.M. Smith$^{4}$} 
{$^{1}$ School of Physics, University of New South Wales, Sydney 2052, Australia\\  
$^{2}$ IGPP, Lawrence Livermore National Laboratory,
L-413, Livermore, CA 94550, USA}
{$^{3}$ Department of Physics, University of California at Davis,  Davis, CA 95616, USA\\
$^{4}$ Dept. of Physics, University of Wales at Cardiff, PO Box 913, Cardiff CF2 3YB, UK}
\begin{moriondabstract}
In a spectroscopic survey of the Fornax cluster to $B_J=17.5$ using
the FLAIR spectrograph on the UK Schmidt Telescope we have discovered
seven new compact dwarf cluster members. These were previously thought
to be giant background spirals.  These new members are among the most
compact, high surface brightness dwarf galaxies known with absolute
magnitudes of $M_B\approx-14$ and scale lengths of $\alpha\approx400$
pc.  One in particular may be the first high (normal) surface
brightness dwarf spiral discovered. Three of the new dwarfs are blue
compact dwarfs (BCDs); their inclusion in the cluster increases the
faint end of the BCD luminosity function by a factor of 2 or more.  We
extended the survey 2.5 mag fainter with the 2dF spectrograph and in
our first field found 7 bright emission line galaxies beyond the
Fornax cluster which were unresolved on the sky survey plates.
Galaxies of this type would be missed in most existing galaxy surveys.
\end{moriondabstract}

\section{Introduction}
It is now well-accepted that selection effects can lead to significant
biases in galaxy surveys, especially leading to detections of only a
limited range of surface brightness. Disney\cite{dis76} and others
showed that it is very difficult to detect low surface brightness
(LSB) galaxies in typical surveys because they do not register
above the background of random sky fluctuations. Similarly, high
surface brightness (HSB) galaxies would not show any extension and
would be confused with stars. Much work has been done on LSB
galaxies\cite{phil98} but the HSB bias has generally been
discounted\cite{all79}.  In this paper we demonstrate that there is a
real bias against HSB galaxies in two specific contexts: in nearby
clusters where HSB dwarf galaxies---although resolved---are confused
with background giant galaxies and also in field galaxy surveys based
on photographic surveys where compact galaxies are unresolved.

\section{Compact dwarf galaxies in the Fornax cluster}

Our study of the Fornax cluster dwarfs was motivated by a desire to
investigate the relationship between the different dwarf types as well
as determining if there were any compact dwarf elliptical (cdE)
galaxies in the cluster. For the former we were particularly
interested to determine if any very compact dwarfs were missing from
the Fornax Cluster Catalog (FCC)\cite{ferg89} as these might represent the
faded remnants of blue compact dwarf (BCD) galaxies\cite{dh91}. We
therefore observed a large number of galaxies of small projected size
that were classified as probable background galaxies in the FCC. The
selection is illustrated in Fig.~1 which plots the parameters of a
large number of stars and galaxies in the Fornax field based on
photographic data (see caption).

\begin{figure}
\centering
\epsfxsize=16cm \epsffile{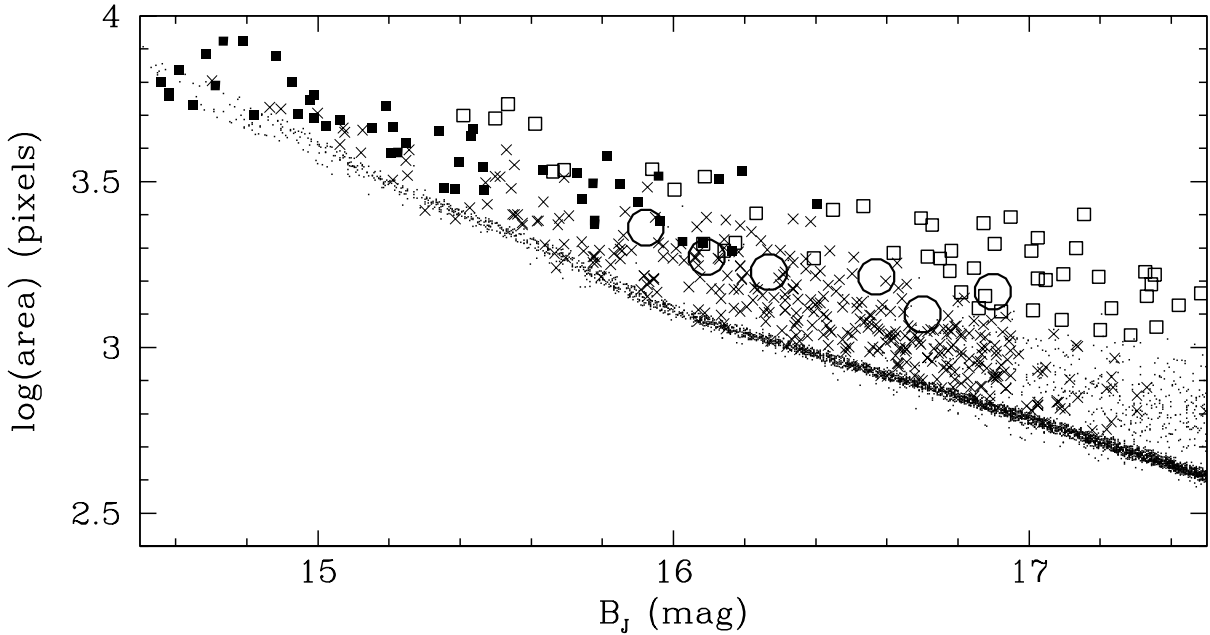}
\centering
\caption{Classification diagram of Fornax galaxies from a sky survey
plate digitized by the automated plate measuring facility (APM). Each
symbol shows the area of an image above the detection threshold as a
function of apparent magnitude.  Galaxies classified in the FCC as
members (confirmed: filled squares; unconfirmed: open squares) are
seen to be less compact than the background galaxies (confirmed:
crosses; unconfirmed: dots). Stars are also plotted as dots forming a
locus with minimum area. The new compact dwarf cluster members are
plotted as large open circles.}
\end{figure}

We observed a large sample of galaxies with small image area with the
FLAIR spectrograph on the UK Schmidt Telescope to determine if they
were cluster members. We obtained redshifts for 453 galaxies, an
almost complete sample of previously unmeasured compact galaxies with
magnitudes brighter than $B=17.3$. We also included 78 galaxies listed
in the FCC as candidate compact dwarf elliptical galaxies in the FCC
to a magnitude limit of $B=17.7$. Our results showed that only one of
the cdE candidates we observed (FCC B2144) was a member, but that its
spectrum was of a blue, actively star-forming galaxy so we identify it
as a BCD not a cdE.

From our data we found 7 new dwarf cluster members that were
classified as probable background galaxies in the FCC.  These are
illustrated in Fig.~2: these are the most compact dwarf galaxies known
in the cluster as is shown by their position in Fig.~1. We have
obtained CCD images of FCC B2144 which emphasize the small size: it
has an absolute magnitude of $M_B=-14.4$ and an exponential scale
length of only $\alpha=0.3$ kpc (for a cluster distance of 15.4
Mpc\cite{bur96}). With a slit spectrum we detected rotation with an
amplitude of around 20 km/s. The spectrum is very blue with strong
emission lines so we classify it as a BCD. At least one of the dwarfs
(FCC B0905) shows evidence of a disk-plus-bulge morphology and we
suggest it may be a true dwarf spiral type, the first discovered.

\begin{figure}
\centering
\epsfxsize=17cm \epsffile{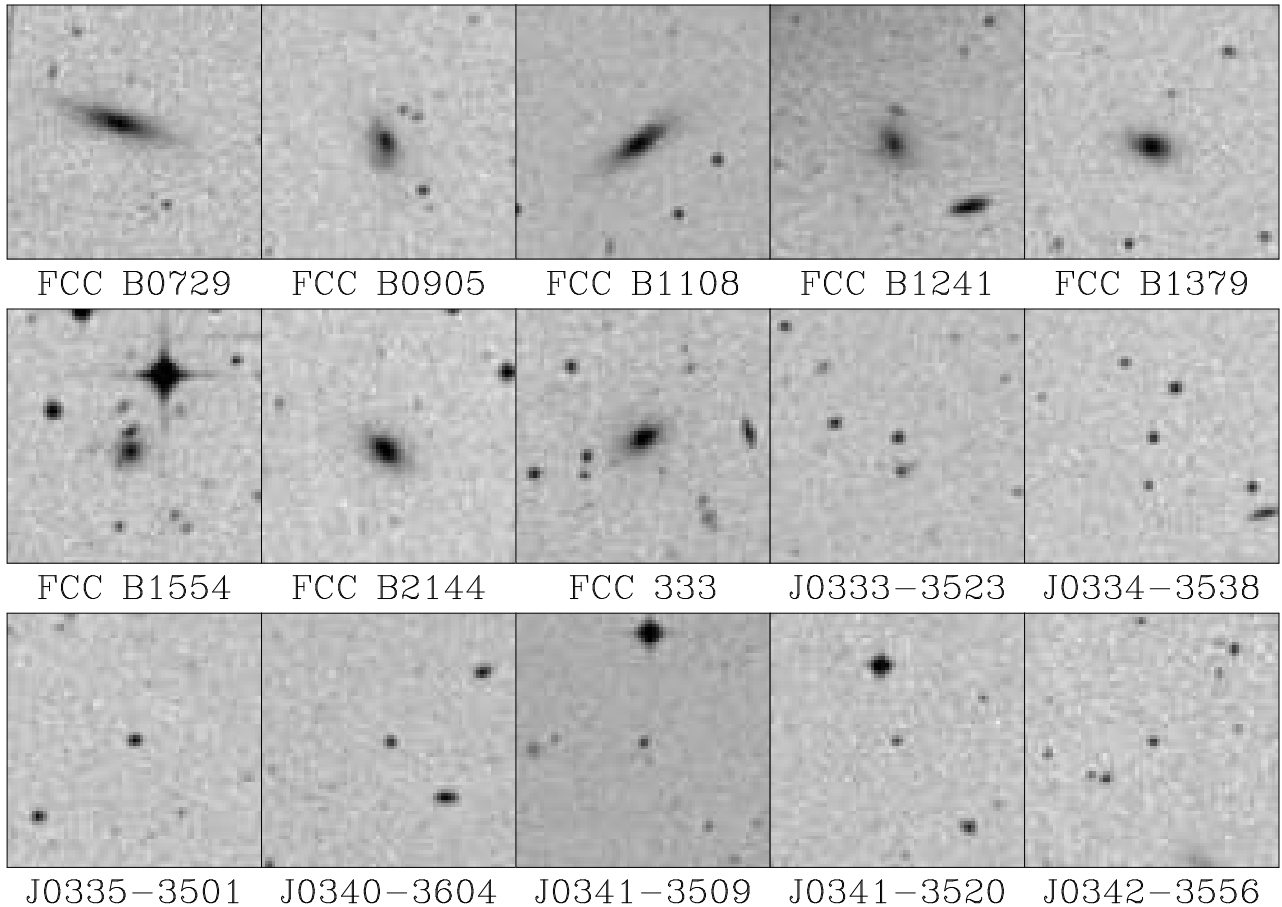}
\centering
\caption{Images of seven new compact dwarfs in the Fornax cluster, a
background giant galaxy (FCC 333: $M_B=-19.5, cz=13600$km/s)
previously thought to be a cluster member and seven new compact field
galaxies we have discovered beyond the cluster ($-17<M_B<-21,
13000<cz<55000$ km/s). J0334$-$3538 is at the same distance as FCC
333, albeit 2 mag fainter.  The images are all from the Digitized Sky
Survey in the $B_J$ band giving a region 2 arcmin across with North to
the top and East to the left.}
\end{figure}

Three of the new cluster dwarfs (the two mentioned above and B1379)
show strong emission lines, so are BCDs. This actually doubles the
number of confirmed BCDs in the cluster and extends their distribution
to much fainter levels. However we prefer to leave the morphological
classifications and instead consider the star-forming dwarf galaxies
as defined by the presence of strong emission lines. We attempt to do
this in Fig.~3 where we compare the magnitudes of the star-forming
dwarfs to all cluster members in the FCC. This shows a strong increase
in the number of star-forming dwarfs to fainter magnitudes although we
prefer not to interpret this in terms of the frequency of
star-formation bursts because we do not have evidence to link the
different dwarfs in one evolutionary sequence\cite{dh91}.

\begin{figure}
\centering
\epsfxsize=16cm \epsffile{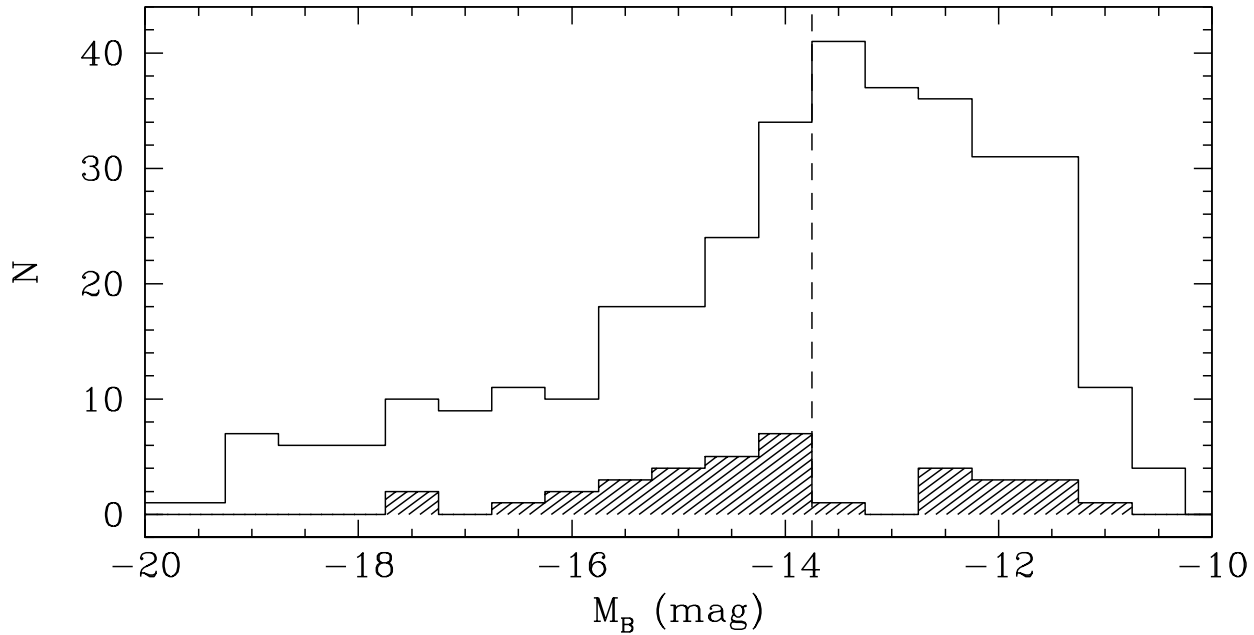}
\centering
\caption{Histogram of the absolute magnitudes of all Fornax cluster
galaxies (upper histogram) and star-forming galaxies (lower, shaded)
in the FCC. The star-forming dwarf galaxies were defined by taking all
late-type dwarfs in the FCC (BCD, Im, Sm, Sdm), adding the new
star-forming dwarfs we found and removing any for which we obtained
spectra with no emission lines. The dashed line gives the limit of
our spectroscopic measurements.}
\end{figure}

\section{Compact giant galaxies in the background}

We also searched for compact dwarf galaxies in the field behind the
cluster sufficiently distant to be unresolved in the photographic
data, i.e. at least 3 times farther away or 2.4 mag fainter, a limit
of about 19.5.  We are using the 2dF spectrograph on the
Anglo-Australian Telescope to identify all objects brighter than
$B=19.7$ in a 12 deg$^2$ region at the centre of the cluster. We
observed some 600 unresolved objects in our first 2dF run: seven of
these were emission-line galaxies and not stars. These are
also shown in Fig.~2. They were all farther away than we projected so
they are not dwarfs, but compact bright galaxies (unless, as suggested
at the meeting, they are dwarfs undergoing particularly intense bursts
of star formation). Whatever these new galaxies turn out to be, they
constitute at least 1-2\% of the local galaxy population but have been
missed by existing surveys based on the photographic sky survey data.

\smallskip\noindent MJD wishes to acknowledge financial support from
the conference organisers and an Australia/France Co-operation in
Astronomy grant.


%
\end{document}